\documentclass{article}

\usepackage{amstext,amssymb,amsmath,amsthm,latexsym}
\usepackage{enumerate,bm}
\usepackage{mathrsfs}
\usepackage[top=1in,bottom=1in,left=1in,right=1in]{geometry}
\usepackage{paralist}
\usepackage{mdwlist}
\usepackage[pdftex,colorlinks,citecolor=blue,linkcolor=blue]{hyperref}

\theoremstyle{plain}
\newtheorem{theorem}{Theorem}[section]

\newtheorem{lemma}[theorem]{Lemma}

\theoremstyle{definition}
\newtheorem{definition}[theorem]{Definition}
\newtheorem{algorithm}[theorem]{Algorithm}
\newtheorem{example}[theorem]{Example}
\newtheorem{notation}[theorem]{Notation}

\theoremstyle{remark}

\DeclareMathOperator*{\lcm}{lcm}
\DeclareMathOperator*{\cmr}{cmr}

\newcommand{\ar}{a}
\newcommand{\pb}{B}
\newcommand{\Bd}{\pb_\dr}
\newcommand{\Be}{\pb_\er}
\newcommand{\Bp}{\pb_\ipr}
\newcommand{\Bq}{\pb_\qr}
\newcommand{\Bs}{\pb_\varepsilon}
\newcommand{\bN}{{\mathbb{N}^n}}
\newcommand{\bM}{\bm{[x]}}
\newcommand{\bQ}{\mathbb{Q}}
\newcommand{\bg}{Buchberger}
\newcommand{\bs}{s}
\newcommand{\bt}{t}
\newcommand{\bzc}{B\'ezout coefficients}
\newcommand{\Cp}{\text{\textsc{cp}}}
\newcommand{\cS}{s}
\newcommand{\cds}{\Omega}
\newcommand{\clm}{\tx^\gamma}
\newcommand{\cn}[1]{N(#1)}
\newcommand{\dd}{d}
\newcommand{\dr}{\textsl{d}}
\newcommand{\dss}{\Lambda}
\newcommand{\ee}{\er_\qr}
\newcommand{\el}{\chi}
\newcommand{\elm}{\el_\qr}
\newcommand{\er}{\textsl{e}}

\newcommand{\fir}{\nu}
\newcommand{\gb}{Gr\"obner}
\newcommand{\IP}{\text{\textsc{ip}}}
\newcommand{\Id}{I_\dr}
\newcommand{\Ip}{I_\ipr}
\newcommand{\Iq}{I_\qr}
\newcommand{\Iu}{I\cap\kux}
\newcommand{\ibr}{b}
\newcommand{\icr}{c}
\newcommand{\idr}{d}
\newcommand{\iod}{\iota_\dr}
\newcommand{\ioe}{\iota_\er}
\newcommand{\iop}{\iota_\ipr}
\newcommand{\ioq}{\iota_\qr}
\newcommand{\ipr}{p}
\newcommand{\iur}{\mu}
\newcommand{\iwr}{w}
\newcommand{\kux}{K[x_1]}
\newcommand{\kxx}{(\kux)[\tx]}
\newcommand{\Lst}[3]{#1_{#2},\dotsc,#1_{#3}}
\newcommand{\lc}{\text{\textrm{lc}}}
\newcommand{\lcc}{\text{\textsc{lc}}}
\newcommand{\lel}{\tf_0}
\newcommand{\lex}{\textsc{lex}}
\newcommand{\lm}{\text{\textrm{lm}}}
\newcommand{\lmc}{\text{\textsc{lm}}}
\newcommand{\lmr}{m}
\newcommand{\lnr}{n}
\newcommand{\lt}{\text{\textrm{lt}}}
\newcommand{\ltc}{\text{\textsc{lt}}}
\newcommand{\mar}{\lambda}
\newcommand{\mas}{\Lambda}
\newcommand{\md}{\sigma_\dr}
\newcommand{\mlt}{\mathrm{mult}}
\newcommand{\mq}{\sigma_\qr}
\newcommand{\mpp}{\sigma_\ipr}
\newcommand{\mr}{m}
\newcommand{\Np}{\mathbb{N}^\ast}
\newcommand{\pd}{\Theta}
\newcommand{\nona}{\kux\setminus K^\ast}
\newcommand{\nonk}{\kux\setminus K}

\newcommand{\Po}[1]{Procedure $\mathcal{#1}$}
\newcommand{\pel}{\chi_\varepsilon}
\newcommand{\pid}[1]{(#1)}

\newcommand{\qa}{\theta}
\newcommand{\qel}{\qa_\qr}
\newcommand{\qr}{q}
\newcommand{\RD}{\kxx\setminus\kux}
\newcommand{\Rd}{R_\dr}
\newcommand{\Rdx}{\Rd [\tx]}
\newcommand{\Rp}{R_\ipr}
\newcommand{\Rq}{R_\qr}
\newcommand{\Rqx}{\Rq [\tx]}
\newcommand{\Rqd}{\Rqx\setminus\Rq}

\newcommand{\rat}{\Rq^\ast\setminus\Rq^\times}
\newcommand{\rhor}{\rho}
\newcommand{\rnd}{(\Rqx)^\ast\setminus\Rq^\times}

\newcommand{\rr}{r}
\newcommand{\rxd}{R[\X]\setminus R}
\newcommand{\sfr}{\omega}

\newcommand{\supp}[1]{\mathrm{supp}(#1)}
\newcommand{\tM}{\bm{[}\tx\bm{]}}
\newcommand{\tas}{F}
\newcommand{\tbs}{G}
\newcommand{\tf}{f}
\newcommand{\tg}{g}
\newcommand{\tx}{\tilde{\X}}
\newcommand{\X}{\bm{x}}
\newcommand{\vr}{v}

\numberwithin{equation}{section}

\begin{document}

\title{A New Type of \gb\ Basis and Its Complexity
\footnotetext{\textit{Email:} \texttt{masm@buaa.edu.cn;~smmath@foxmail.com}}
\footnotetext{\textit{Address:} School of Mathematical Sciences, Beihang University, Beijing 100191, China.}
 \footnotetext{\textit{Key
words and phrases:} \gb\ bases, polynomial ideal, ideal bases, eliminant, zero divisor, modular method, pseudo-division, intermediate coefficient swell problem, intermediate expression swell problem.}}
\author{Sheng-Ming Ma}
\date{}

\maketitle

\begin{abstract}
The new type of ideal basis introduced herein constitutes a compromise between the \gb\ bases based on the \bg's algorithm and the characteristic sets based on the Wu's method.
It reduces the complexity of the traditional \gb\ bases and subdues the notorious intermediate expression swell problem and intermediate coefficient swell problem to a substantial extent.
The computation of an $S$-polynomial for the new bases requires at most $O(m\ln^2m\ln\ln m)$ word operations whereas $O(m^6\ln^2m)$ word operations are requisite in the \bg's algorithm.
Here $m$ denotes the upper bound for the numbers of terms both in the leading coefficients and for the rest of the polynomials.
The new bases are for zero-dimensional polynomial ideals and based on univariate pseudo-divisions.
However in contrast to the pseudo-divisions in the Wu's method for the characteristic sets, the new bases retain the algebraic information of the original ideal and in particular, solve the ideal membership problem.
In order to determine the authentic factors of the eliminant, we analyze the multipliers of the pseudo-divisions and develop an algorithm over principal quotient rings with zero divisors.
\end{abstract}


\section{Introduction}

The theory of \gb\ bases \cite{AL94,BW93,CLO15,DL06,EH12,GG13,GP08,KR00} has been established as a standard tool in algebraic geometry and computer algebra and solves many significant problems in mathematics, science and engineering \cite{BW98}.
Nonetheless the computational complexity of \gb\ bases often demands an enormous amount of computing time and storage space even for problems of moderate sizes \cite[P116]{CLO15} \cite[P616]{GG13}.
The striking phenomena include the \emph{intermediate coefficient swell problem} in the computation of \gb\ bases over the rational field with respect to the \lex\ ordering, as well as the \emph{intermediate expression swell problem} referring to a generation of a huge number of intermediate polynomials during the implementation of the algorithm.
These challenges stimulate decades of ardent endeavors in developing various methodologies such as the normal selection strategies and signatures \cite{BW93,Buc85,EF17,Fau02,GMN91,SW10,SW11,GVW16}, the modular and $p$-adic techniques and Hensel lifting \cite{Arn03,Ebe83,Gra93,Pau92,ST89,Tra89,Win87}, as well as the \gb\ basis conversion methods like the FGLM algorithm \cite{FGLM93} and \gb\ Walk \cite{CKM97,CLO05,Stu95}.
However albeit with all these endeavors over the decades, the high-level complexity associated with the \gb\ basis computations remains a conundrum.

The Wu's method \cite{Wu83} is based on pseudo-divisions and thus more efficient than the method of \gb\ bases.
However the pseudo-divisions only yield the zero locus or radical ideal of the original ideal and hence lose too much algebraic information to solve algebraic problems like the \emph{ideal membership problem}.

The new type of \gb\ basis introduced herein is a compromise between the \gb\ bases and the characteristic sets based on the Wu's method.
We take the variable of the eliminant as the parametric variable and use univariate pseudo-divisions to reduce the computational complexity and retain the algebraic information of the original ideal simultaneously.

In Algorithm \ref{Algo:PseudoEliminant} we compute the pseudo-eliminant $\pel$ and pseudo-basis $\Bs$ of the original ideal.
Then we compare $\pel$ with the multipliers of the pseudo-divisions to discriminate its compatible and incompatible parts in Definition \ref{Def:CompatibleDivisors}.
Theorem \ref{Thm:CompatiblePart} establishes that the compatible part of $\pel$ constitutes a bona fide factor of the eliminant $\el$.

We conduct a complete analysis of the incompatible part $\IP (\pel)$ of $\pel$ based on modular algorithms whose moduli are the composite divisors of $\IP (\pel)$ as in Definition \ref{Def:CompositeDivisor}.
The principal quotient rings (PQR) thus obtained might contain zero divisors and we have to redefine the $S$-polynomials in Definition \ref{Def:SpolynPQR} carefully.
In Algorithm \ref{Algo:ProperEliminant} we obtain the proper eliminants and proper bases by proper divisions as in Theorem \ref{Thm:ProperReduction}.
We prove rigorously in Theorem \ref{Thm:IncompatiblePart} that the nontrivial proper divisors as in Definition \ref{Def:ProperFactor} that are obtained in Algorithm \ref{Algo:ProperEliminant} are the bone fide factors of the eliminant $\el$ of the original ideal.

The characterizations of the new type of basis $\Bs$, $\Bq$ and $\Bp$ are in \eqref{LeadTermMod}, \eqref{ExactLeadTermChar} and \eqref{ProperLeadTermChar} respectively.
This new type of basis in \eqref{NewBases} corresponds to a decomposition of the original ideal in \eqref{IdealDecomposition}.

A special scenario consisting of two basis elements in Lemma \ref{Lemma:OldGBasis} reveals that the \bg's algorithm contains the Extended Euclidean Algorithm computing the greatest common divisor of their leading coefficients and in particular, the \bzc\ that might swell to an enormous size.
This might help to unveil the mystery of the high-level complexity associated with the traditional \gb\ basis computations such as the intermediate coefficient and expression swell problems.
By contrast the computation of our new type of $S$-polynomial in \eqref{NewSPolynComput} yields the above results in one step without the \bzc\ in a conspicuously moderate number of requisite word operations.
Example \ref{Expl:FullModularAlgo} contains a specific example of the new type of basis.

For a ring $R$ we denote $R^\ast:=R\setminus\{0\}$ and use $R^\times$ to denote the set of units in $R^\ast$.

\section{A Pseudo-division Algorithm over PIDs}\label{Sec:DivisionAlgm}

Let $R$ be a PID and $R[\X]$ a polynomial algebra over $R$.
Let us denote the set of monomials in $\X=(\Lst x1n)$ as $\bM:=\{\X^\alpha\colon\alpha\in\bN\}$ with a monomial ordering denoted as $\succ$.
A nonzero ideal $I\subset R[\X]$ is called a \emph{monomial} ideal if $I$ is generated by monomials in $\bM$.

\begin{notation}\label{Notation:LeadingEntities}
Let $f=\sum_\alpha c_\alpha\X^\alpha$ be a polynomial in $R[\X]$.
We denote the \emph{support} of $f$ as $\supp f:=\{\X^\alpha\in\bM\colon c_\alpha\ne 0\}$.
In particular, we define $\supp f:=\{1\}$ when $f\in R^\ast$ and $\supp f:=\emptyset$ when $f=0$.

Hereafter we use the following terminologies.
The \emph{leading term} of $f$ is a term $c_\beta\X^\beta$ that satisfies $\X^\beta:=\max_{\succ}\{\X^\alpha\in\supp f\}$ and is denoted as $\ltc (f):=c_\beta\X^\beta$.
Here $\max_{\succ}$ denotes the maximal element with respect to the monomial ordering $\succ$.
The \emph{leading monomial} of $f$ is the monomial $\X^\beta$ and denoted as $\lmc (f):=\X^\beta$.
The \emph{leading coefficient} of $f$ is the coefficient $c_\beta$ and denoted as $\lcc (f):=c_\beta\in R^\ast$.

Let $\pb=\{\ibr_j\colon 1\le j\le s\}$ be a polynomial set in $R[\X]\setminus\{0\}$.
We denote the leading monomial set $\{\lmc (\ibr_j)\colon 1\le j\le s\}$ as $\lmc (\pb)$.
Let us also denote the monomial ideal generated by $\lmc (\pb)$ in $R[\X]$ as $\langle\lmc (\pb)\rangle$.

In what follows we use $\gcd (a,b)$ and $\lcm (a,b)$ to denote the greatest common divisor and least common multiple of $a,b\in R^\ast$ respectively.
\end{notation}

\begin{definition}[Term pseudo-reduction over a PID $R$]\label{Def:TermReduction}
\hfill

For $f\in\rxd$ and $g\in R[\X]\setminus\{0\}$, suppose that $f$ has a term $c_\alpha\X^\alpha$ such that $\X^\alpha\in\supp f\cap\langle\lmc (g)\rangle$.
Then we can make a \emph{pseudo-reduction} of the term $c_\alpha\X^\alpha$ by $g$ as follows.
\begin{equation}\label{TermReduction}
h=\iur f-\frac{\lmr\X^\alpha}{\ltc (g)}g
\end{equation}
with the multipliers $\lmr:=\lcm (c_\alpha,\lcc (g))$ and $\iur:=\lmr/c_\alpha\in R^\ast$.
We call $h$ the \emph{remainder} of the pseudo-reduction and $\iur$ the \emph{interim multiplier} on $f$ with respect to $g$.
\end{definition}

\begin{definition}[Pseudo-reduced polynomial]
\hfill

A polynomial $\rr\in R[\X]$ is \emph{pseudo-reduced} with respect to a polynomial set $\pb=\{\ibr_j\colon 1\le j\le s\}\subset\rxd$ if $\supp\rr\cap\langle\lmc (\pb)\rangle=\emptyset$.
In particular, this includes the special case when $\rr=0$ and hence $\supp\rr=\emptyset$.
We also say that $\rr$ is \emph{pseudo-reducible} with respect to $\pb$ if it is not pseudo-reduced with respect to $\pb$, i.e., $\supp\rr\cap\langle\lmc (\pb)\rangle\ne\emptyset$.
\end{definition}

\begin{theorem}[Pseudo-division over a PID $R$]\label{Thm:PseudoReduction}
\hfill

Suppose that $\pb=\{\ibr_j\colon 1\le j\le s\}\subset\rxd$ is a polynomial set.
For every $f\in R[\X]$, there exist a multiplier $\mar\in R^\ast$, a remainder $\rr\in R[\X]$ and quotients $\qr_j\in R[\X]$ for $1\le j\le s$ such that
\begin{equation}\label{PseudoDivisionExpression}
\mar f=\sum_{j=1}^s\qr_j\ibr_j+\rr,
\end{equation}
where $\rr$ is pseudo-reduced with respect to $\pb$.
Moreover, the polynomials in \eqref{PseudoDivisionExpression} satisfy the following condition:
\begin{equation}\label{DivisionCond}
\lmc (f)=\max\bigl\{\max_{1\le j\le s}\{\lmc (\qr_j\ibr_j)\},\lmc (\rr)\bigr\}.
\end{equation}
\end{theorem}
\begin{proof}
If $f$ is not pseudo-reduced with respect to $\pb$, we define $\X^\alpha:=\max_{\succ}\{\supp f\cap\langle\lmc (\pb)\rangle\}$.
There exists some $j$ such that $\X^\alpha$ is divisible by $\lmc (\ibr_j)$.
Let us make a pseudo-reduction of the term $c_\alpha\X^\alpha$ of $f$ by $\ibr_j$ as in \eqref{TermReduction}.
We denote the remainder as $h$ and it is easy to see that $\X^\alpha\succ\X^\beta:=\max_{\succ}\{\supp h\cap\langle\lmc (\pb)\rangle\}$.
Such term pseudo-reductions terminate in finite steps until the remainder $h$ is pseudo-reduced with respect to $\pb$ since the monomial ordering $\succ$ is a well-ordering.
Hence follows the representation \eqref{PseudoDivisionExpression} in which the multiplier $\mar\in R^\ast$ is a product of such interim multipliers $\iur$ as in \eqref{TermReduction}.

To prove the equality \eqref{DivisionCond}, it suffices to prove that it holds for the term pseudo-reduction in \eqref{TermReduction}.
\end{proof}

We call the expression in \eqref{PseudoDivisionExpression} a \emph{pseudo-division} of $f$ by $\pb$.
More specifically, we name the polynomial $r$ in \eqref{PseudoDivisionExpression} as a \emph{remainder} of $f$ and $\mar\in R^\ast$ in \eqref{PseudoDivisionExpression} as a \emph{multiplier} of the pseudo-division.
We say that $f$ \emph{pseudo-reduces} to the \emph{remainder} $r$ via the \emph{multiplier} $\mar\in R^\ast$ with respect to $\pb$.

\section{Pseudo-eliminants of Zero-dimensional Ideals}\label{Sec:PseudoGroebner}

In this section we consider the case when the PID $R$ in Section \ref{Sec:DivisionAlgm} satisfies $R=\kux$ with $K$ being a field and $x_1$ the least variable of $\X$.
We always treat the algebra $K[\X]$ as the algebra $\kxx$ with the variables $\tx:=(\Lst x2n)$.
With $\alpha=(\Lst\alpha 2n)$, we denote a monomial $x_2^{\alpha_2}\cdots x_n^{\alpha_n}$ as $\tx^\alpha$.
Hence $\lcc (f)\in (\kux)^\ast$ for $f\in\kxx$.
Let us use $\pid g$ to denote the principal ideal in $\kux$ that is generated by $g\in\kux$.
Recall that $\langle f\rangle$ denotes a principal ideal in $\kxx$ that is generated by $f\in\kxx$.

In what follows let us suppose that $I$ is a zero-dimensional ideal of $K[\X]=\kxx$.

\begin{definition}[Eliminant]\label{Def:Eliminant}
\hfill

For a zero-dimensional ideal $I\subset\kxx$, we denote the generator of the principal ideal $\Iu$ as $\el$ such that $\Iu=\pid\el$.
We call $\el$ the \emph{eliminant} of the zero-dimensional ideal $I$ henceforth.
\end{definition}

\begin{definition}[$S$-polynomial]\label{Def:SPolynomial}
\hfill

Suppose that $f,g\in\RD$.
Let us denote $\lmr:=\lcm (\lcc (f),\lcc (g))\in (\kux)^\ast$ and $\clm:=\lcm (\lmc (f),\lmc (g))\in\tM$.
Then the polynomial
\begin{equation}\label{SPolynomialDef}
S(f,g):=\frac{\lmr\clm}{\ltc (f)}f-\frac{\lmr\clm}{\ltc (g)}g
\end{equation}
is called the \emph{$S$-polynomial} of $f$ and $g$.
\end{definition}

When $g\in (\kux)^\ast$ and $f\in\RD$, we take $\lmc (g)=1$ and $\lmr=\lcm (\lcc (f),g)$.
The $S$-polynomial in \eqref{SPolynomialDef} is now defined as:
\begin{equation}\label{SpecialSPoly}
S(f,g):=\frac\lmr{\lcc (f)}f-\lmr\cdot\lmc (f).
\end{equation}

By the identity $\lmr/g=\lcc (f)/\idr$ with $\idr:=\gcd (\lcc (f),g)\in (\kux)^\ast$, we can easily deduce Lemma \ref{Lemma:UnnecessaryConst} as follows.
The same for the proof of Lemma \ref{Lemma:RelativePrimePairs}.

\begin{lemma}\label{Lemma:UnnecessaryConst}
When $g\in (\kux)^\ast$ and $f\in\RD$, the $S$-polynomial in \eqref{SpecialSPoly} satisfies:
\begin{equation}\label{SpecialSPolyEssence}
\idr S(f,g)=(f-\ltc (f))\cdot g:=f_1g
\end{equation}
with $f_1:=f-\ltc (f)$.
\end{lemma}

\begin{lemma}\label{Lemma:RelativePrimePairs}
For $f,g\in\RD$, suppose that $\lmc (f)$ and $\lmc (g)$ are relatively prime.
Let us denote $\idr:=\gcd (\lcc (f),\lcc (g))$.
Then their $S$-polynomial in \eqref{SPolynomialDef} satisfies:
\begin{equation}\label{CoprimeReduction}
\idr S(f,g)=f_1g-g_1f=f_1\cdot\ltc (g)-g_1\cdot\ltc (f)
\end{equation}
with $f_1:=f-\ltc (f)$ and $g_1:=g-\ltc (g)$.
Moreover, we have:
\begin{equation}\label{LeadTwoMonomial}
\lmc (S(f,g))=\max\{\lmc (f_1g),\lmc (g_1f)\}.
\end{equation}
\end{lemma}

\begin{lemma}\label{Lemma:TriangleIdentity}
If $\lcm (\lmc (f),\lmc (g))\in\langle\lmc (h)\rangle$ for $f,g,h\in\RD$, then we have the following triangular relationship among their $S$-polynomials:
\begin{equation}\label{TriangleIdentity}
\mar S(f,g)=\frac{\mar\cdot\lcm (\ltc (f),\ltc (g))}{\lcm (\ltc (f),\ltc (h))}S(f,h)-\frac{\mar\cdot\lcm (\ltc (f),\ltc (g))}{\lcm (\ltc (g),\ltc (h))}S(g,h),
\end{equation}
where the multiplier $\mar:=\lcc (h)/\idr$ with $\idr:=\gcd (\lmr,\lcc (h))\in\kux$ and $\lmr:=\lcm (\lcc (f),\lcc (g))$.
Henceforth let us also call the identity \eqref{TriangleIdentity} the \emph{triangular identity} of $S(f,g)$ with respect to $h$.
\end{lemma}
\begin{proof}
It suffices to write the numerator $\lmr\clm$ in the definition of $S$-polynomial in \eqref{SPolynomialDef} into $\lmr\clm=\lcm (\ltc (f),\ltc (g))$.
In fact, the identity \eqref{TriangleIdentity} readily follows if we also write the $S$-polynomials $S(f,h)$ and $S(g,h)$ into this form.
\end{proof}

\begin{algorithm}[Pseudo-eliminant of a zero-dimensional ideal over $\kux$]\label{Algo:PseudoEliminant}
\hfill

Input: A finite polynomial set $\tas\subset\kxx\setminus K$.

Output: A pseudo-eliminant $\pel\in (\kux)^\ast$, pseudo-basis $\Bs\subset\langle\tas\rangle\setminus\kux$ and multiplier set $\mas\subset\nonk$.

Initialization: A temporary basis set $\tbs:=\tas\setminus\kux$; a multiplier set $\mas:=\emptyset$; a temporary set $\mathfrak{S}:=\emptyset$ of $S$-polynomials.
We initialize $\lel:=\gcd (\tas\cap\kux)$ or $\lel:=0$ depending on $\tas\cap\kux\ne\emptyset$ or not.

For each pair $\tf,\tg\in\tbs$ with $\tf\ne\tg$, we invoke \Po Q as follows to compute their $S$-polynomial $S(f,g)$.

\Po Q:

\leftskip=5mm 
\begin{itshape}
Input: $f,g\in\RD$.

If $\lmc (\tf)$ and $\lmc (\tg)$ are relatively prime, we define $\idr:=\gcd (\lcc (f),\lcc (g))$ as in \eqref{CoprimeReduction}.
If $\idr\in\nonk$, we add $\idr$ into the multiplier set $\mas$, and we do nothing otherwise.
Then we disregard the $S$-polynomial $S(\tf,\tg)$.

If $\lcm (\lmc (f),\lmc (g))\in\langle\lmc (h)\rangle$ for an $h\in\tbs\setminus\{\tf,\tg\}$, and the triangular identity \eqref{TriangleIdentity} has never been applied to the same triplet $\{f,g,h\}$ before, we compute the multiplier $\mar$ as in \eqref{TriangleIdentity}.
If $\mar\in\nonk$, we add $\mar$ into the multiplier set $\mas$, and we do nothing otherwise.
Then we disregard the $S$-polynomial $S(\tf,\tg)$.

If neither of the above two cases is true, we compute their $S$-polynomial $S(\tf,\tg)$ as in \eqref{SPolynomialDef}.
Then we add $S(\tf,\tg)$ into the set $\mathfrak{S}$.
\end{itshape}

End of $\mathcal{Q}$
\leftskip=0mm

We recursively repeat \Po P as follows for the pseudo-reductions of all the $S$-polynomials in the set $\mathfrak{S}$.

\Po P:

\leftskip=5mm 
\begin{itshape}
For an $S\in\mathfrak{S}$, we invoke Theorem \ref{Thm:PseudoReduction} to make a pseudo-reduction of $S$ by the temporary basis set $\tbs$.

If the multiplier $\mar\in\nonk$ in \eqref{PseudoDivisionExpression}, we add $\mar$ into the multiplier set $\mas$.

If the remainder $\rr=0$, we do nothing and continue with the algorithm.

If the remainder $\rr\in\RD$, we add $\rr$ into $\tbs$.
For every $\tf\in\tbs\setminus\{\rr\}$, we invoke \Po Q to compute the $S$-polynomial $S(\tf,\rr)$.

If the remainder $\rr\in\nonk$, we redefine $\lel:=\gcd (\rr,\lel)$.

Then we delete $S$ from the set $\mathfrak{S}$.
\end{itshape}

End of $\mathcal{P}$
\leftskip=0mm

Finally we define $\pel:=\lel$ and $\Bs:=\tbs$ respectively.

\Po R:

\leftskip=5mm 
\begin{itshape}
For every $\tf\in\Bs$, if $\idr:=\gcd (\lcc (f),\pel)\in\nonk$, we add $\idr$ into the multiplier set $\mas$.
\end{itshape}

End of $\mathcal{R}$
\leftskip=0mm

We output $\pel$, $\Bs$ and $\mas$.
\qed\end{algorithm}

\begin{definition}[Pseudo-eliminant $\pel$; pseudo-basis $\Bs$; multiplier set $\mas$]\label{Def:PseudoBasis}
\hfill

Henceforth we call the univariate polynomial $\pel$ obtained via Algorithm \ref{Algo:PseudoEliminant} a \emph{pseudo-eliminant} of the zero-dimensional ideal $I$.
We also call the basis set $\Bs$ a \emph{pseudo-basis} of the ideal $I$ and $\mas$ its \emph{multiplier set}.
\end{definition}

\begin{lemma}\label{Lemma:PseudoEliminantTerminate}
Algorithm \ref{Algo:PseudoEliminant} terminates in finite steps.
\end{lemma}
\begin{proof}
The termination of the algorithm follows from $\kxx$ being Noetherian.
In fact, in \Po P of Algorithm \ref{Algo:PseudoEliminant}, the monomial ideal $\langle\lmc (\tbs)\rangle$ is strictly expanded each time we add the remainder $r\in\RD$ into $\tbs$ since $r$ is pseudo-reduced with respect to $\tbs\setminus\{r\}$.
\end{proof}

\section{Pseudo-eliminant Divisors and Compatibility}\label{Section:PseudoEliminantDivisors}

In this section we prove that the compatible part $\Cp (\pel)$ of the pseudo-eliminant $\pel$ is a bona fide factor of the eliminant $\el$.

\begin{definition}[Compatible and incompatible divisors and parts]\label{Def:CompatibleDivisors}
\hfill

For an irreducible factor $\ipr$ of $\pel$ with multiplicity $i$, if $\ipr$ is relatively prime to every multiplier $\mar$ in $\mas$, then $\ipr^i$ is called a \emph{compatible divisor} of $\pel$.
Otherwise $\ipr^i$ is called an \emph{incompatible divisor} of $\pel$.

We name the product of all the compatible divisors of $\pel$ as the \emph{compatible part} of $\pel$ and denote it as $\Cp (\pel)$.
The \emph{incompatible part} of $\pel$ is defined as $\IP (\pel):=\pel/\Cp (\pel)$.
\end{definition}

\begin{algorithm}[Compatible part $\Cp (\pel)$ and squarefree decomposition of the incompatible part $\IP (\pel)$]\label{Algo:CompatiblePartPseudoEliminant}
\hfill

Input: The pseudo-eliminant $\pel\in (\kux)^\ast$ and multiplier set $\mas\subset (\kux)^\ast$ that are obtained from Algorithm \ref{Algo:PseudoEliminant}.

Output: Compatible part $\Cp (\pel)$ and a squarefree decomposition $\{\cds_i\colon 1\le i\le s\}$ of the incompatible part $\IP (\pel)$.

We initialize $\cds_i=\emptyset$ for $1\le i\le s$ and make a squarefree factorization of the pseudo-eliminant $\pel$ as $\pel=\prod_{i=1}^s q_i^i$.

For every $\mar\in\mas$, we compute $\idr_{\mar i}:=\gcd (\mar,q_i)$.
If $\idr_{\mar i}\in\nonk$, we check whether $\idr_{\mar i}$ is relatively prime to every element $\sfr$ that is already in $\cds_i$.
If not, we substitute $\idr_{\mar i}$ by $\idr_{\mar i}/\gcd (\idr_{\mar i},\sfr)$.
We also substitute the $\sfr$ in $\cds_i$ by both $\gcd (\idr_{\mar i},\sfr)$ and $\sfr/\gcd (\idr_{\mar i},\sfr)$ if neither of them is in $K^\ast$.
Let us repeat the process until either $\idr_{\mar i}\in K^\ast$, or $\idr_{\mar i}\in\nonk$ is relatively prime to every element in $\cds_i$.
Then we add $\idr_{\mar i}$ into $\cds_i$ if $\idr_{\mar i}\in\nonk$.

Finally, we output $\pel/\prod_{i=1}^s\prod_{\sfr\in\cds_i}\sfr^i$ as the compatible part $\Cp (\pel)$.
We also output $\{\cds_i\colon 1\le i\le s\}$ as a squarefree decomposition of the incompatible part $\IP (\pel)$.
\qed\end{algorithm}

\begin{definition}[Composite divisor $\sfr^i$]\label{Def:CompositeDivisor}
\hfill

For an element $\sfr$ of the univariate polynomial set $\cds_i$ for $1\le i\le s$ obtained in Algorithm \ref{Algo:CompatiblePartPseudoEliminant}, we call its $i$-th power $\sfr^i$ a \emph{composite divisor} of the incompatible part $\IP (\pel)$.
\end{definition}

\begin{lemma}\label{Lemma:SyzygyTransform}
Suppose that each $f_j$ in $\tas:=\{f_j\colon 1\le j\le s\}\subset\RD$ has the same leading monomial $\lmc (f_j)=\tx^\alpha\in\tM$.
If $f=\sum_{j=1}^s f_j$ satisfies $\lmc (f)\prec\tx^\alpha$, then
there exist multipliers $\ibr,\ibr_j\in (\kux)^\ast$ for $1\le j<s$ such that
\begin{equation}\label{SPolynomialExpansion}
\ibr f=\sum_{1\le j<s}\ibr_jS(f_j,f_s)
\end{equation}
with the $S$-polynomial $S(f_j,f_s)$ being defined as in \eqref{SPolynomialDef}.
Moreover, for each irreducible polynomial $\ipr\in\nonk$, we can always relabel the subscripts of the polynomial set $\tas$ such that the multiplier $\ibr$ of $f$ in \eqref{SPolynomialExpansion} is not divisible by $\ipr$.
\end{lemma}
\begin{proof}
Let us denote $l_j:=\lcc (f_j)$ for $1\le j\le s$ and $\lmr_j:=\lcm (l_j,l_s)$ for $1\le j<s$.
From $\lmc (f)\prec\tx^\alpha$ we can deduce that $\sum_{j=1}^s l_j=0$.
Now the identity in \eqref{SPolynomialExpansion} can be easily corroborated if we define the multipliers as follows:
\begin{equation}\label{SyzygyMultipliers}
\ibr:=\lcm_{1\le j<s}\Bigl(\frac{\lmr_j}{l_j}\Bigr);~\ibr_j:=\frac{\ibr l_j}{\lmr_j}~(1\le j<s).
\end{equation}

Let us denote the multiplicity of $\ipr$ in $l_j$ as $\mlt_\ipr (l_j)\ge 0$.
We relabel the subscripts of $f_j$ and $l_j$ for $1\le j\le s$ such that $\mlt_\ipr (l_s)=\min_{1\le j\le s}\{\mlt_\ipr (l_j)\}$.
Then $\mlt_\ipr (\lmr_j/l_j)=\mlt_\ipr (l_s/\gcd (l_j,l_s))=0$ for $1\le j<s$.
Thus the multiplier $\ibr$ in \eqref{SyzygyMultipliers} is not divisible by $\ipr$.
\end{proof}

\begin{theorem}\label{Thm:CompatiblePart}
Let $\el$ and $\pel$ be the eliminant and pseudo-eliminant of a zero-dimensional ideal $I\subset\kxx$ respectively.
Then $\el$ is divisible by the compatible divisors of $\pel$ and hence by the compatible part $\Cp (\pel)$ of $\pel$.
\end{theorem}
\begin{proof}
Let $\ipr^i$ be a compatible divisor of the pseudo-eliminant $\pel$ as in Definition \ref{Def:CompatibleDivisors}.
Let us prove that the eliminant $\el$ is also divisible by $\ipr^i$.

Let $\widetilde\tas:=\tbs\cup\{\lel\}:=\{f_j\colon 0\le j\le s\}\subset\kxx\setminus K$ be the basis of the ideal $I$ after the Initialization in Algorithm \ref{Algo:PseudoEliminant} with $\lel\in\nona$.
The eliminant $\el\in I\cap\kux$ can be written as $\el=\sum_{j=0}^s h_jf_j$ with $h_j\in\kxx$ for $0\le j\le s$.
Let us denote $\tx^\beta:=\max_{0\le j\le s}\{\lmc (h_jf_j)\}$.
Then we collect and rename the elements in the set $\{f_j\in\widetilde\tas\colon\lmc (h_jf_j)=\tx^\beta,0\le j\le s\}$ into a new set $\pb_t:=\{g_j\colon 1\le j\le t\}$.
And the subscripts of the functions $\{h_j\}$ are adjusted accordingly.
In this way we have:
\begin{equation}\label{RevisedRepresentation}
\el=\sum_{j=1}^th_jg_j+\sum_{f_i\in\widetilde\tas\setminus\pb_t}h_if_i
\end{equation}

If we denote $\ltc (h_j):=c_j\tx^{\alpha_j}$ with $c_j\in (\kux)^\ast$ for $1\le j\le t$, then according to Lemma \ref{Lemma:SyzygyTransform}, there exist multipliers $\ibr,\ibr_j\in (\kux)^\ast$ for $1\le j<t$ such that the polynomial $\tg:=\sum_{j=1}^t\ltc (h_j)\cdot g_j$ satisfies the following identity:
\begin{equation}\label{LeadingMonomialSyzygy}
\ibr\tg=\sum_{1\le j<t}\ibr_jS(c_j\tx^{\alpha_j}g_j,c_t\tx^{\alpha_t}g_t).
\end{equation}
Moreover, we can relabel the subscript set in \eqref{LeadingMonomialSyzygy} such that $\mlt_\ipr (\ibr)=0$ by Lemma \ref{Lemma:SyzygyTransform}.

In the case of $\pb_t\subset\RD$, if we denote $\tx^{\gamma_j}:=\lcm (\lmc (g_j),\lmc (g_t))$, then we can simplify the $S$-polynomials in \eqref{LeadingMonomialSyzygy} as follows:
\begin{equation}\label{SPolynomialRelation}
S(c_j\tx^{\alpha_j}g_j,c_t\tx^{\alpha_t}g_t)=\lmr_j\tx^{\beta-\gamma_j}S(g_j,g_t)
\end{equation}
with $\lmr_j:=\lcm (c_j\cdot\lcc (g_j),c_t\cdot\lcc (g_t))/\lcm (\lcc (g_j),\lcc (g_t))$ for $1\le j<t$.

Let $\Bs=\{\widetilde\tg_k\colon 1\le k\le\tau\}\subset\RD$ be the pseudo-basis of the ideal $I$ obtained in Algorithm \ref{Algo:PseudoEliminant}, in which we have pseudo-reduced every $S$-polynomial $S(g_j,g_t)$ in \eqref{SPolynomialRelation} by $\Bs$.
More specifically, as per Theorem \ref{Thm:PseudoReduction}, there exist a multiplier $\mar_j\in (\kux)^\ast$ as well as a remainder $\rhor_j\pel$ with $\rhor_j\in\kux$ and quotients $q_{jk}\in\kxx$ for $1\le k\le\tau$ such that the following pseudo-reduction holds for $1\le j<t$:
\begin{equation}\label{SPolynomialReduction}
\mar_jS(g_j,g_t)=\sum_{k=1}^\tau q_{jk}\widetilde\tg_k+\rhor_j\pel
\end{equation}
with $\mlt_\ipr (\mar_j)=0$ for $1\le j<t$ since $\ipr^i$ is a compatible divisor.
As per \eqref{DivisionCond}, it readily follows that for $1\le j<t$:
\begin{equation}\label{OrderSPolynomialReduction}
\max_{1\le k\le\tau}\{\lmc (q_{jk}\widetilde\tg_k)\}=\lmc (S(\tg_j,\tg_t))\prec\tx^{\gamma_j}.
\end{equation}

Based on a combination of \eqref{SPolynomialRelation} and \eqref{SPolynomialReduction}, it is straightforward to obtain a pseudo-reduction of the $S$-polynomial $S(c_j\tx^{\alpha_j}g_j,c_t\tx^{\alpha_t}g_t)$ in \eqref{LeadingMonomialSyzygy} by the pseudo-basis $\Bs$ with the same multiplier $\mar_j$.
This combined with \eqref{LeadingMonomialSyzygy} yield the following representation:
\begin{equation}\label{TempRepresent}
\ibr\mar\tg=\sum_{k=1}^\tau q_k\widetilde\tg_k+\eta\pel,
\end{equation}
with $\eta,q_k\in\kxx$ for $1\le k\le\tau$.
The multiplier $\ibr\mar$ is relatively prime to the compatible divisor $\ipr^i$.
Moreover, from \eqref{SPolynomialRelation} and \eqref{OrderSPolynomialReduction} we have the following inequality for \eqref{TempRepresent}:
\begin{equation}\label{DecreaseLeadingMonomial}
\max\bigl\{\max_{1\le k\le\tau}\{q_k\widetilde\tg_k\},\eta\pel\bigr\}\prec\tx^\beta.
\end{equation}
Now we can rewrite the representation in \eqref{RevisedRepresentation} into a new one in terms of $\Bs$ and $\pel$ as follows.
\begin{equation}\label{FinalRepresentation}
\ibr\mar\el=\sum_{k=1}^\tau \iur_k\widetilde\tg_k+\iur_0\pel
\end{equation}
with $\iur_k\in\kxx$ for $0\le k\le\tau$.
And the leading monomials in \eqref{FinalRepresentation} satisfy
\begin{equation}\label{RepresentationLeadingMonomial}
\max\bigl\{\max_{1\le k\le\tau}\{\lmc (\iur_k\widetilde\tg_k)\},\lmc (\iur_0)\bigr\}\prec\tx^\beta
\end{equation}
according to \eqref{DecreaseLeadingMonomial}.

In summary, the leading monomials in the representation \eqref{FinalRepresentation} strictly decrease from those in the representation \eqref{RevisedRepresentation}, up to the multiplier $\ibr\mar$ that satisfies $\mlt_\ipr (\ibr\mar)=0$.
In particular, when $\lel$ satisfies $\lel\in\pb_t$ as in \eqref{RevisedRepresentation}, we can prove by \eqref{SpecialSPoly} that the conclusion is still sound.

We repeat the above procedure of rewriting the representations of the eliminant $\el$ so as to strictly reduce the orderings of their leading monomials.
Moreover, the multipliers for the representations are always relatively prime to the compatible divisor $\ipr^i$.
Since the monomial ordering is a well-ordering, the above process halts after a finite number of repetitions.
In this way we shall reach a representation bearing the following form:
\begin{equation}\label{Finally}
\fir\el=h\pel,
\end{equation}
where the multiplier $h\in (\kux)^\ast$.
In particular, the multiplier $\fir\in (\kux)^\ast$ is relatively prime to the compatible divisor $\ipr^i$.
Hence follows the conclusion.
\end{proof}

\section{Analysis of Incompatible Divisors via Modular Method}
\label{Sec:IncompatibleModular}

Let $K$ be a field and $\qr\in\nonk$.
With $R=\kux$, the quotient ring $R/\pid\qr$ is called a \emph{Principal ideal Quotient Ring} and abbreviated as a PQR henceforth.
Consider the set $\Rq:=\{r\in\kux\colon\deg (r)<\deg (\qr)\}$ with $\deg (r)=0$ for $r\in K$.
We redefine the two binary operations, the addition and multiplication, on $\Rq$ such that it is isomorphic to the PQR $R/\pid\qr$.
We call $\Rq$ a \emph{normal} PQR and define an epimorphism $\mq\colon R\rightarrow\Rq$ as $\mq (f):=r$ via the division $f=h\qr+r$ with the quotient $h\in\kux$ and unique remainder $r\in\Rq$.
We can also define an injection $\ioq\colon\Rq\hookrightarrow R$ as $\ioq (r):=r$ since $\Rq\subset\kux$.
The epimorphism $\mq$ can be extended to $\mq\colon\kxx\rightarrow\Rqx$ that is the identity map on the variables $\tx$.
Similarly the injection $\ioq$ can be extended to $\ioq\colon\Rqx\rightarrow\kxx$.

\begin{definition}[Term reduction in \text{$\Rqx$}]\label{Def:TermReductionPQR}
\hfill

For $f\in\Rqd$ and $g\in\rnd$, suppose that $f$ has a term $\icr_\alpha\tx^\alpha$ with $\tx^\alpha\in\supp f\cap\langle\lmc (g)\rangle$.
We define the multipliers $\iur:=\mq (\lcm (l_\alpha,l_g)/l_\alpha)$ and $\lmr:=\mq (\lcm (l_\alpha,l_g)/l_g)$ with $l_\alpha:=\ioq (\icr_\alpha)$ and $l_g:=\ioq (\lcc (g))$.
We can make a \emph{reduction} of the term $\icr_\alpha\tx^\alpha$ by $g$ as follows.
\begin{equation}\label{TermReductionPQR}
h=\iur f-\frac{\lmr\tx^\alpha}{\lmc (g)}g.
\end{equation}
We call $h$ the \emph{remainder} of the reduction and $\iur$ the \emph{interim multiplier} on $f$ with respect to $g$.
\end{definition}

\begin{definition}[Properly reduced polynomial]\label{Def:ProperlyReduced}
\hfill

A nonzero term $\icr_\alpha\tx^\alpha\in\Rqx$ is said to be \emph{properly reducible} with respect to $\tas=\{\Lst f1s\}\subset\Rqd$ if there exists an $f_j\in\tas$ such that $\tx^\alpha\in\langle\lmc (f_j)\rangle$ and the interim multiplier $\iur$ with respect to $f_j$ as in \eqref{TermReductionPQR} satisfies $\iur\in\Rq^\times$.
We say that a polynomial $f\in\Rqx$ is \emph{properly reduced} with respect to $\tas$ if none of its terms is properly reducible with respect to $\tas$.
\end{definition}

The proof of the following theorem is almost a verbatim repetition of that for Theorem \ref{Thm:PseudoReduction}.

\begin{theorem}[Proper division or reduction]\label{Thm:ProperReduction}
\hfill

Suppose that $\tas=\{\Lst f1s\}$ are polynomials in $\Rqd$.
For every $f\in\Rqx$, there exist a multiplier $\mar\in\Rq^\times$ as well as a remainder $\rr\in\Rqx$ and quotients $q_j\in\Rqx$ for $1\le j\le s$ such that:
\begin{equation}\label{ProperDivisionExpression}
\mar f=\sum_{j=1}^s q_jf_j+\rr,
\end{equation}
where $\rr$ is properly reduced with respect to $\tas$.
Moreover, the polynomials in \eqref{ProperDivisionExpression} satisfy the following condition:
\begin{equation}\label{ProperDivisionCond}
\lmc (f)=\max\{\max_{1\le j\le s}\{\lmc (q_j)\cdot\lmc (f_j)\},\lmc (r)\}.
\end{equation}
\end{theorem}

\begin{definition}[$S$-polynomial over $\Rq$]\label{Def:SpolynPQR}
\hfill

Suppose that $f\in\Rqd$ and $g\in\rnd$.
Let us denote $l_f:=\ioq(\lcc (f))$ and $l_g:=\ioq(\lcc (g))$ in $(\kux)^\ast$ respectively.
We also define the multipliers $\lmr_f:=\mq (\lcm (l_f,l_g)/l_f)$ and $\lmr_g:=\mq (\lcm (l_f,l_g)/l_g)$ as well as the monomial $\clm:=\lcm (\lmc (f),\lmc (g))\in\tM$.
Then the following polynomial:
\begin{equation}\label{SPolyPQR}
S(f,g):=\frac{\lmr_f\clm}{\lmc (f)}f-\frac{\lmr_g\clm}{\lmc (g)}g
\end{equation}
is called the \emph{$S$-polynomial} of $f$ and $g$ in $\Rqx$.
\end{definition}

In particular, when $f\in\Rqd$ and $\tg\in\rat$, we can take $\lmc (\tg)=1$ and then the $S$-polynomial in \eqref{SPolyPQR} bears the following form:
\begin{equation}\label{IdentitySpecialSPoly}
S(f,\tg):=\lmr_ff-\lmr_\tg\tg\cdot\lmc (f)=\mq\Bigl(\frac {l_\tg}\idr\Bigr)(f-\ltc (f))
\end{equation}
with $\idr:=\gcd (l_\tf,l_\tg)$ and $l_\tg:=\ioq (\tg)$.

When $\lcc (f)\in\rat$ for $f\in\Rqd$, there is another special kind of $S$-polynomial
\begin{equation}\label{BeheadSPolyPQR}
S(f,\qr):=\lnr_ff=\lnr_f (f-\ltc (f))
\end{equation}
with $\lnr_f:=\mq (\lcm (l_f,\qr)/l_f)$.

We can easily deduce the following lemma.
\begin{lemma}\label{Lemma:RelativePrimePQR}
For $f,g\in\Rqd$, suppose that $\lmc (f)$ and $\lmc (g)$ are relatively prime.
With the same notations as in Definition \ref{Def:SpolynPQR}, let us also denote $\idr:=\gcd (l_f,l_g)$.
Then their $S$-polynomial satisfies:
\begin{equation}\label{CoprimeReductionPQR}
\mq (\idr)\cdot S(f,g)=f_1\cdot\ltc (g)-g_1\cdot\ltc (f)=f_1g-g_1f
\end{equation}
with $f_1:=f-\ltc (f)$ and $g_1:=g-\ltc (g)$.
\end{lemma}

Let us use the same notations as in Definition \ref{Def:SpolynPQR}.
For $f,g\in\rnd$ without both of them in $\rat$, we define $\cmr (g\vert f):=\lmr_f\clm/\lmc (f)$.
Then the $S$-polynomial $S(f,g)=\cmr (g\vert f)\cdot f-\cmr (f\vert g)\cdot g$, by which we can deduce the following lemma.

\begin{lemma}\label{Lemma:TriangleIdentityPQR}
For $f,g,h\in\rnd$ with at most one of them in $\rat$, if $\lcm (\lmc (f),\lmc (g))\in\langle\lmc (h)\rangle$, then we have the following relationship between their $S$-polynomials:
\begin{equation}\label{TriangleIdentityPQR}
\mar S(f,g)=\frac{\mar\cdot\cmr (g\vert f)}{\cmr (h\vert f)}S(f,h)-\frac{\mar\cdot\cmr (f\vert g)}{\cmr (h\vert g)}S(g,h).
\end{equation}
Here the multiplier $\mar:=\mq (l_h/\idr)\in\Rq^\ast$ with $l_h:=\ioq (\lcc (h))$ and $\idr:=\gcd (\lcm (l_f,l_g),l_h)$.
\end{lemma}

For a multiplicity $i$ satisfying $1\le i\le s$ and composite divisor $\sfr^i$ with $\sfr\in\cds_i$ as in Definition \ref{Def:CompositeDivisor}, let us denote $\sfr^i$ as the modulus $\qr$ and consider the normal PQR $\Rq$ with $R=\kux$.
Suppose that we have a unique factorization $\qr=\sfr^i=u\prod_{k=1}^tp_k^i$ with $t\in\Np$ and $u\in R^\times$.
When $t>1$ the irreducible factors $\{p_k\colon 1\le k\le t\}\subset R^\ast\setminus R^\times$ are pairwise relatively prime.
Then every $a\in\Rq^\ast$ has a \emph{standard} representation as follows:
\begin{equation}\label{StdRepsPQR}
a\sim a^\ast:=\prod_{k=1}^t p_k^{\beta_k},\quad 0\le\beta_k\le i;
\qquad a=a^\times\cdot a^\ast.
\end{equation}

\begin{algorithm}[Proper eliminant and proper basis over a normal PQR $\Rq$]\label{Algo:ProperEliminant}
\hfill

Input: A finite polynomial set $\tas\subset\Rqd$.

Output: A proper eliminant $\ee\in\Rq$ and proper basis $\Bq\subset\Rqd$.

Initialization: A temporary set $\mathfrak{S}:=\emptyset$ in $\Rqx\setminus\Rq$ for $S$-polynomials; a temporary $e\in\Rq$ as $e:=0$.

For each pair $\tf,\tg\in\tas$ with $\tf\ne\tg$, we invoke \Po R as follows to compute their $S$-polynomial $S(f,g)$.

\Po R:

\leftskip=5mm 
\begin{itshape}
If $\lmc (\tf)$ and $\lmc (\tg)$ are relatively prime, we compute the multiplier $\mq (\idr)$ as in \eqref{CoprimeReductionPQR} with $\idr:=\gcd (\ioq (\lcc (f)),\ioq (\lcc (g)))$.
If $\mq (\idr)\in\rat$, we compute the $S$-polynomial $S(\tf,\tg)$ as in \eqref{CoprimeReductionPQR} and then add it into the set $\mathfrak{S}$.
If $\mq (\idr)\in\Rq^\times$, we disregard $S(\tf,\tg)$.

If $\lcm (\lmc (f),\lmc (g))\in\langle\lmc (h)\rangle$ for an $h\in\tas\setminus\{\tf,\tg\}$, and the triangular identity \eqref{TriangleIdentityPQR} has not been applied to the same triplet $\{f,g,h\}$ before, we compute the multiplier $\mar$ as in \eqref{TriangleIdentityPQR}.
If $\mar\in\rat$, we compute the $S$-polynomial $S(\tf,\tg)$ as in \eqref{SPolyPQR} and then add it into the set $\mathfrak{S}$.
If $\mar\in\Rq^\times$, we disregard $S(\tf,\tg)$.

If neither of the above two cases is true, we compute the $S$-polynomial $S(\tf,\tg)$ as in \eqref{SPolyPQR} and then add it into the set $\mathfrak{S}$.
\end{itshape}

End of $\mathcal{R}$
\leftskip=0mm

We recursively repeat \Po P as follows for proper reductions of all the $S$-polynomials in $\mathfrak{S}$.

\Po P:

\leftskip=5mm 
\begin{itshape}
For an $S\in\mathfrak{S}$, we invoke Theorem \ref{Thm:ProperReduction} to make a proper reduction of $S$ by $\tas$.

If the remainder $\rr=0$, we do nothing and continue with the algorithm.

If the remainder $\rr\in\Rq^\times$, we halt the algorithm and output $\ee=1$.

If the remainder $\rr\in\Rqd$, we add $\rr$ into $\tas$.
For every $\tf\in\tas\setminus\{\rr\}$, we invoke \Po R to compute the $S$-polynomial $S(\tf,\rr)$.

If the remainder $\rr\in\rat$ and $e=0$, we redefine $\er:=\mq (\gcd (\ioq (\rr),q))$.

If the remainder $\rr\in\rat$ and $\er\in\Rq^\ast$, we compute $\idr=\mq (\gcd (\ioq (\rr),\ioq (\er)))$.
If $\idr\notin\pid\er\subset\Rq$, we redefine $\er:=\idr$.

Then we delete $S$ from $\mathfrak{S}$.
\end{itshape}

End of $\mathcal{P}$
\leftskip=0mm

Next we recursively repeat \Po Q as follows for proper reductions of the special kinds of $S$-polynomials in \eqref{IdentitySpecialSPoly} and \eqref{BeheadSPolyPQR}.

\Po Q:

\leftskip=5mm 
\begin{itshape}
If $\mathfrak{S}=\emptyset$ and $\er=0$, then for every $\tf\in\tas$ with $\lcc (\tf)\in\rat$, we compute the $S$-polynomial $S(\tf,\qr)$ as in \eqref{BeheadSPolyPQR} and add it into $\mathfrak{S}$ if this has not been done for $\tf$ in a previous step.

Then we recursively repeat \Po P.

If $\mathfrak{S}=\emptyset$ and $\er\in\Rq^\ast$, then for every $\tf\in\tas$  with $\lcc (\tf)\in\rat$, if $\mq (\idr)\in\rat$ with $\idr:=\gcd (\ioq (\lcc (\tf)),\ioq (\er))$, we compute the $S$-polynomial $S(\tf,\er)$ as in \eqref{IdentitySpecialSPoly} and add it into $\mathfrak{S}$ unless one of its associates had been added into $\mathfrak{S}$ in a previous step.

Then we recursively repeat \Po P.
\end{itshape}

End of $\mathcal{Q}$
\leftskip=0mm

Finally we define and output $\ee:=\er$ and $\Bq:=\tas$ respectively.
\qed\end{algorithm}

\begin{definition}[Proper eliminant $\ee$; proper basis $\Bq$; modular eliminant $\elm$]\label{Def:ProperEliminant}
\hfill

With the ideal $\Iq:=\mq (I)$, we call the standard representation $\ee^\ast$ as in \eqref{StdRepsPQR} of $\ee\in\Iq\cap\Rq$ obtained in Algorithm \ref{Algo:ProperEliminant}, whether it is zero or not, a \emph{proper} eliminant of $\Iq$.
In what follows let us simply denote $\ee:=\ee^\ast$.
We also call the final polynomial set $\Bq$ obtained in Algorithm \ref{Algo:ProperEliminant} a \emph{proper} basis of $\Iq$.
Moreover, $\elm:=\mq (\el)$ is called the \emph{modular} eliminant of $\Iq$.
\end{definition}

\begin{lemma}\label{Lemma:nPQRSyzygy}
Let $\tas=\{f_j\colon 1\le j\le s\}\subset\Rqd$ be a polynomial set.
Suppose that for $1\le j\le s$, each $f_j$ has the same leading monomial $\lmc (f_j)=\tx^\alpha$.
If $f=\sum_{j=1}^s f_j$ satisfies $\lmc (f)\prec\tx^\alpha$, then
there exist multipliers $\ibr,\ibr_j\in\Rq^\ast$ for $1\le j<s$ such that
\begin{equation}\label{SPolynomialExpansionPQR}
\ibr f=\sum_{1\le j<s}\ibr_jS(f_j,f_s)
\end{equation}
with the $S$-polynomial $S(f_j,f_s)$ being defined as in \eqref{SPolyPQR}.
Moreover, for every irreducible factor $\ipr$ of the composite divisor $\qr$, we can always relabel the subscripts of the polynomial set $\tas=\{f_j\colon 1\le j\le s\}$ such that the multiplier $\ibr\in\Rq^\ast$ in \eqref{SPolynomialExpansionPQR} is not divisible by $\ipr$.
\end{lemma}
\begin{proof}
Let us denote $l_j:=\ioq (\lcc (f_j))$ for $1\le j\le s$.
We define the multipliers $\lmr_j:=\lcm (l_j,l_s)/l_j$ for $1\le j<s$.
Let us also define a multiplier $\ibr:=\mq (\ar)$ with $\ar:=\lcm_{1\le j<s}(\lmr_j)$.
The identity \eqref{SPolynomialExpansionPQR} can be corroborated by the multipliers $\ibr_j:=\mq (\ar_j)$ with $\ar_j:=\ar/\lmr_j$ for $1\le j<s$.
Moreover, for an irreducible factor $\ipr$ of the composite divisor $\qr$, we change the order of the elements in $\tas$ so that $\mlt_\ipr (l_s)=\min_{1\le j\le s}\{\mlt_\ipr (l_j)\}$.
Hence $\mlt_\ipr (\lmr_j)=0$ for $1\le j<s$.
And thus $\mlt_\ipr (\ar)=0$.
\end{proof}

\begin{theorem}\label{Thm:IncompatiblePart}
Let $\qr=\sfr^i$ be a composite divisor and $\ee$ and $\elm$ denote the proper and modular eliminants respectively as in Definition \ref{Def:ProperEliminant}.
\begin{enumerate}
\item\label{item:ProperZeroRemainder} If the proper eliminant $\ee=0$, the eliminant $\el$ is divisible by the composite divisor $\qr=\sfr^i$ and the modular eliminant $\elm=0$.

\item\label{item:ProperNonZeroRemainder} If the proper eliminant $\ee\in\Rq^\ast$, the eliminant $\el$ is divisible by $\ioq (\ee)$ and the modular eliminant $\elm^\ast=\ee$.
\end{enumerate}
\end{theorem}
\begin{proof}
The proof is similar to that of Theorem \ref{Thm:CompatiblePart}.
After fixing an irreducible factor $\ipr$ of the composite divisor $\qr$, we repeatedly rewrite the representations of the modular eliminant $\elm$ using Lemma \ref{Lemma:nPQRSyzygy} and the proper reductions of $S$-polynomials in Algorithm \ref{Algo:ProperEliminant}.
In this way we strictly reduce the orderings of the leading monomials of the representations.
Moreover, the multipliers for the representations are always relatively prime to the irreducible factor $\ipr$.
Since the monomial ordering is a well-ordering, the above process halts after a finite number of repetitions.
In this way we shall obtain a representation of $\elm$ bearing the following form:
\begin{equation}\label{FinallyPQR}
\fir\elm=h\ee
\end{equation}
with $h\in\Rq$ and in particular, the multiplier $\fir\in\Rq^\ast$ not divisible by the irreducible factor $\ipr$ of the composite divisor $\qr$.
The conclusion follows from an analysis of \eqref{FinallyPQR}.
\end{proof}

\section{A New Type of Basis for Zero-dimensional Ideals}\label{Sec:InductiveGroebner}

\begin{definition}[Proper divisors $\qel$]\label{Def:ProperFactor}
\hfill

For every composite divisor $\qr=\sfr^i$, there corresponds to a proper eliminant $\ee$ as in Definition \ref{Def:ProperEliminant}.
We define a \emph{proper divisor} $\qel\in\kux$ in accordance with $\ee$ as follows.
If $\ee\in\Rq^\times$, we define $\qel:=1$;
If $\ee=0$, we define $\qel:=\qr$;
If $\ee\in\rat$, we define $\qel:=\ioq (\ee)$.
\end{definition}

The following conclusion is straightforward.

\begin{theorem}\label{Thm:Eliminant}
The eliminant $\el$ is the product of the compatible part $\Cp (\pel)$ and all the proper divisors $\qel$.
\end{theorem}

With an almost verbatim repetition of the proof for Theorem \ref{Thm:CompatiblePart}, we can prove the following conclusion.

\begin{lemma}\label{Lemma:PseudoReps}
Let $\Bs=\{g_k\colon 1\le k\le\tau\}$ be a pseudo-basis of a zero-dimensional ideal $I$ and $\Cp (\pel)$ the compatible part of the pseudo-eliminant $\pel$ associated with $\Bs$.
For every $f\in I$, there exist $\{\vr_k\colon 0\le k\le\tau\}\subset\kxx$ and a multiplier $\mar$ relatively prime to $\Cp (\pel)$ such that:
\begin{equation}\label{PseudoReps}
\mar f=\sum_{k=1}^\tau \vr_k\tg_k+\vr_0\pel.
\end{equation}
Moreover, we have:
\begin{equation}\label{PseudoRepCond}
\lmc (f)=\max\bigl\{\max_{1\le k\le\tau}\{\lmc (\vr_k\tg_k)\},\lmc (\vr_0)\bigr\}.
\end{equation}
\end{lemma}

\begin{lemma}\label{Lemma:IdealMembership}
Let us treat the compatible part $\Cp (\pel)$ as the modulus $\idr$ and define the normal PQR $\Rd$ and the epimorphism $\md\colon\kxx\rightarrow\Rdx$ as before.
Then for $\Id:=\md (I)$ and $\Bd:=\md (\Bs)$, we have an ideal identity in $\Rdx$ as follows.
\begin{equation}\label{LeadTermMod}
\langle\ltc (\Id)\rangle=\langle\ltc (\Bd)\rangle.
\end{equation}
\end{lemma}
\begin{proof}
For every $\tg\in \Id$, there exists $\tf\in I$ such that $\md (\tf)=\tg$ and $\md (\lcc (\tf ))=\lcc (\tg)\in\Rd^\ast$.
Both \eqref{PseudoReps} and \eqref{PseudoRepCond} hold for $\tf$.
We apply $\md$ to the identity \eqref{PseudoReps} and collect the subscript $k$ into a set $\dss$ if $\lmc (\vr_k)\cdot\lmc (g_k)=\lmc (\tf)$ and $\md (\lcc (\vr_kg_k))=\md (\lcc (\vr_k)\cdot\lcc (g_k))\in\Rd^\ast$.
We have $\dss\ne\emptyset$ as per \eqref{PseudoRepCond}.
Hence follows the identity \eqref{LeadTermMod}.
\end{proof}

We have the following conclusions similar to Lemma \ref{Lemma:PseudoReps} and Lemma \ref{Lemma:IdealMembership} whose proofs are omitted.

\begin{lemma}\label{Lemma:ProperReps}
Let $\qr$ be a composite divisor and $\ee$ and $\Bq=\{\tg_k\colon 1\le k\le\tau\}$ be the proper eliminant and proper basis of $\Iq=\mq (I)$ respectively.
For every $f\in\Iq$, there exist a multiplier $\mar\in\Rq^\times$ and $\{\vr_k\colon 0\le k\le\tau\}\subset\Rqx$ such that:
\begin{equation}\label{ProperReps}
\mar f=\sum_{k=1}^\tau \vr_k\tg_k+\vr_0\ee.
\end{equation}
Moreover, we have:
\begin{equation}\label{ProperRepCond}
\lmc (f)=\max\bigl\{\max_{1\le k\le\tau}\{\lmc (\vr_k)\cdot\lmc (\tg_k)\},\lmc (\vr_0\ee)\bigr\}.
\end{equation}
In particular, the above conclusions are still sound when the proper eliminant $\ee=0$.
\end{lemma}

\begin{lemma}
Let $\qr$ be a composite divisor and $\Iq=\mq (I)$.
Let $\ee\in\Rq\setminus\Rq^\times$ and $\Bq$ denote the proper eliminant and proper basis of $\Iq$ obtained in Algorithm \ref{Algo:ProperEliminant} respectively.
If $\ee=0$, then we have:
\begin{equation}\label{ExactLeadTermChar}
\langle\ltc (\Iq)\rangle=\langle\ltc (\Bq)\rangle.
\end{equation}
If $\ee\in\rat$, let us treat $\ee$ as the modulus $\ipr$ and define the normal PQR $\Rp$.
We also define $\Ip:=\mpp (I)$ and $\Bp:=\mpp (\Bs)$.
Then we have:
\begin{equation}\label{ProperLeadTermChar}
\langle\ltc (\Ip)\rangle=\langle\ltc (\Bp)\rangle.
\end{equation}
\end{lemma}

In summary, we have the following new type of basis for a zero-dimensional ideal.

\begin{theorem}
Let $\dr=\Cp (\pel)$ be the compatible part and $\pd$ the set of nontrivial proper divisors in Definition \ref{Def:ProperFactor}.
Then we have the following decomposition of a zero-dimensional ideal $I$.
\begin{equation}\label{IdealDecomposition}
I=(I+\langle\dr\rangle)\cap\bigcap_{\qel\in\pd}(I+\langle\qel\rangle).
\end{equation}
We have a new type of basis in accordance with the above ideal decomposition:
\begin{equation}\label{NewBases}
(\iod (\Bd)\cup\{\dr\})\cup\bigcup_{\er\in\pd}(\ioe (\Be)\cup\{\er\}),
\end{equation}
where $\dr$ and $\iod (\Bd)$ are as in \eqref{LeadTermMod}.
Here $\er=\qel\in\pd$ and $\ioe (\Be)$ denotes either $\ioq (\Bq)$ in \eqref{ExactLeadTermChar} or $\iop (\Bp)$ in \eqref{ProperLeadTermChar}.
\end{theorem}

\section{Complexity Comparison and Example}
\label{Sec:ComplexityComparison}

Recall that the traditional $S$-polynomial of $f,g\in K[\X]\setminus\{0\}$ over a field $K$ is defined as $\cS (f,g):=\X^\eta(f/\lt (f)-g/\lt (g))$ with $\X^\eta:=\lcm (\lm (f),\lm (g))$.
Here $\lt (f)$ denotes the leading term of $f$ over $K$ and the same for $\lt (g)$.

\begin{lemma}\label{Lemma:OldGBasis}
Suppose that $I=\langle f,g\rangle$ is a zero-dimensional ideal in $\kxx$ such that $\ltc (f)=a\tx^\alpha$ and $\ltc (g)=b\tx^\beta$ with $a,b\in (\kux)^\ast$.
The \bg's algorithm for the traditional \gb\ bases computes the $S$-polynomial $S(f,g)$ in \eqref{SPolynomialDef} for the new type of basis.
In essence it implements the Extended Euclidean Algorithm to compute $\gcd (a',b')$ with $a':=a/\lc (a)$ and $b':=b/\lc (b)$.
Moreover, the \bzc\ of $\gcd (a',b')$ are the coefficient factors of $S(f,g)$.
\end{lemma}
\begin{proof}
We compute the $S$-polynomial $S(f,g)$ as in \eqref{SPolynomialDef} for the new type of basis over $\kux$ as follows.
\begin{equation}\label{NewSPolynComput}
S(f,g)=\lambda\tx^{\gamma-\alpha}f_1-\mu\tx^{\gamma-\beta}g_1,
\end{equation}
where $f_1:=f-\ltc (f)$ and $g_1:=g-\ltc (g)$.
Here $\tx^\gamma:=\lcm (\tx^\alpha,\tx^\beta)$.
And the multipliers $\lambda:=\mr/a=b/\rho$ and $\mu:=\mr/b=a/\rho$ with $\mr:=\lcm (a,b)$ and $\rho:=\gcd (a,b)$.

With $\deg (a)\ge\deg (b)$, the traditional $S$-polynomial $\cS (f,g)$ and its further reductions in the \bg's algorithm are equivalent to the polynomial division of $a'$ by $b'$.
If $\cS (f,g)$ is reduced to $h$ with $\ltc (h)=r\tx^\gamma$, then $r$ is exactly the remainder of the division.
The procedure of adding $h$ into the basis $\{f,g\}$ and reducing the traditional $S$-polynomial $\cS (g,h)$ is equivalent to a polynomial division of $b'$ by $r/\lc (r)$.
We can continue to show that the \bg's algorithm essentially implements the Extended Euclidean Algorithm to compute both $\rho:=\gcd (a',b')$ and its \bzc\ $\bs$ and $\bt$, which yields $\iwr:=\rho\tx^\gamma+\bs\tx^{\gamma-\alpha}f_1/\lc (a)+\bt\tx^{\gamma-\beta}g_1/\lc (b)$.
A reduction of the traditional $S$-polynomials $\cS (f,\iwr)$ and $\cS (g,\iwr)$ by $\iwr$ leads to $tS(f,g)/(\lc (a)\lc (b))$ and $-sS(f,g)/(\lc (a)\lc (b))$ respectively.
\end{proof}

We obtained the $S$-polynomial in \eqref{NewSPolynComput} in one step without the \bzc.
This substantially scales down the number of intermediate polynomials for the \emph{intermediate expression swell problem} and the sizes of the intermediate coefficients for the \emph{intermediate coefficient swell problem}.

Since the worst-case complexity associated with \bg's algorithm is still an open problem, we do not address the problem here.
Instead, a meticulous complexity analysis shows that the number of requisite word operations for the computation of the $S$-polynomial $S(f,g)$ in \eqref{NewSPolynComput} is $O(m\ln^2m\ln\ln m)$ whereas that of \bg's reduction process in the proof of Lemma \ref{Lemma:OldGBasis} is $O(m^6\ln^2m)$.
Here $m:=\max\{d,N\}$ with $\dd:=\max\{\deg (\lcc (f)),\deg (\lcc (g))\}$.
And $N=\max\{\cn{f_1},\cn{g_1}\}$ for $f_1=f-\ltc (f)$ and $g_1=g-\ltc (g)$ with $\cn{f_1}$ denoting the number of elements in $\supp{f_1}$ and the same for $\cn{g_1}$.

\begin{example}\label{Expl:FullModularAlgo}
Suppose the ideal $I=\langle f,g,h\rangle\subset\bQ [x,y,z]$ with
\begin{equation*}
\begin{aligned}
f&=-z^2(z+1)^3x+y;\quad g=z^4(z+1)^6x-y^2;\\
h&=-x^2y+y^3+z^4(z-1)^5.
\end{aligned}
\end{equation*}

The eliminant $\el$ bears the following form:
\begin{equation*}
\begin{aligned}
\el=z^6(z&-1)^5(z^{13}+9z^{12}+36z^{11}+84z^{10}+126z^9+126z^8\\
&+85z^7+31z^6+19z^5-9z^4+4z^3-4z^2-3z-1).
\end{aligned}
\end{equation*}

The multiplier set $\mas=\{z^2(z+1)^3,~z^4(z+1)^6-1\}$.
The new type of basis is as follows.
With the modulus $\ipr=z^6$ and over the normal PQR $\Rp\simeq K[z]/\pid{z^6}$, the modular basis $\Bp$ of $\Ip$ bears the form:
\[
\Bp\left\{\begin{aligned}
b_1&:=z^2(z+1)^3y;&b_2&:=y^2;\\
b_3&:=z^2(z+1)^3x-y;&b_4&:=x^2y-z^4(z-1)^5.
\end{aligned}\right.
\]
With the modulus being the compatible part $\qr=\Cp (\pel)=\el/z^6$ and over the normal PQR $\Rq\simeq K[z]/\pid{q}$, the modular basis $\Bq$ of $\Iq$ bears the form:
\begin{equation*}
\Bq\biggl\{
\begin{aligned}
a_1&:=z^2(z+1)^3(z^4(z+1)^6-1)y+z^6(z+1)^3(z-1)^5;\\
a_2&:=z^4(z+1)^6(z^4(z+1)^6-1)x+z^6(z+1)^3(z-1)^5.
\end{aligned}
\biggr.
\end{equation*}

We list the traditional reduced \gb\ basis of $I$ in the Appendix for comparison.
\end{example}

A direction for future research is to generalize the new type of basis to ideals of positive dimensions as well as to enhance its computational efficiency.
A complexity analysis that is parallel to those in \cite{Dub90,KM96,MM82,MM84,MR13,May89} on the traditional \gb\ bases shall be interesting.
Some inherent connections have been found between the \gb\ bases and characteristic sets \cite{Laz92,WDM20,Wan16}.
We are curious whether the new type of basis can shed some new light on these connections.

\bigskip

\noindent{\bf Acknowledgement.}
The author would like to express his gratitude to Bruno \bg\ for his encouragement and advice for further research.
The author would also like to gratefully acknowledge the generous help from Shaoshi Chen during the preparation of the paper and the encouragement from Dongming Wang.

\newpage
\section*{Appendix}

\begin{align*}
g_1&=20253807z^2y+264174124z^{23}+1185923612z^{22}+850814520z^{21}-3776379304z^{20}-6824277548z^{19}\\
&+1862876196 z^{18}+12815317453z^{17}+3550475421z^{16}+2124010584z^{15}-35582561480z^{14}\\
&+42918431554z^{13}-41728834070z^{12}+35649844325z^{11}-17049238505z^{10}+3388659963z^9\\
&+930240431z^8-61146095z^7-518331181z^6.\\
g_2&=20253807y^2+903303104z^{23}+4102316224z^{22}+3140448384z^{21}-12683487983z^{20}\\
&-23996669428z^{19}+4804720290z^{18}+43739947868z^{17}+14906482335z^{16}+9051639768z^{15}\\
&-121400613331z^{14}+139970660534z^{13}-138071007235z^{12}+118589702914z^{11}-55199680030z^{10}\\
&+11927452134z^9+2021069107z^8-38017822z^7-1768266833z^6;\\
g_3&=2592487296z^2x+(7777461888z-2592487296)y+108083949263z^{23}+486376518055z^{22}\\
&+349557551130z^{21}-1558206505718z^{20}-2820179010211z^{19}+788268739077z^{18}\\
&+5350420983851z^{17}+1476923019345z^{16}+689330555757z^{15}-14602936038043z^{14}\\
&+17386123487861z^{13}-16350039201517z^{12}+13787524468420z^{11}-6235683207154z^{10}\\
&+786997920594z^9+628350552934z^8-64382649769z^7-206531133875z^6;\\
g_4&=20253807x^2y+1037047036z^{23}+4686773132z^{22}+3455561112z^{21}-14868243976z^{20}\\
&-27470438972z^{19}+6731446644z^{18}+51651585868z^{17}+16267315284z^{16}+7429467573z^{15}\\
&-141636109619z^{14}+163168836472z^{13}-155454190640z^{12}+135706468958z^{11}-62903516282z^{10}\\
&+11263865469z^9+2500312823z^8+197272975z^7-1682438629z^6-101269035z^5+20253807z^4.
\end{align*}

\bibliographystyle{plain}

\end{document}